# Towards the maximum efficiency design of a perovskite solar cell by material properties tuning: A multidimensional approach

Manfred G. Kratzenberg[1,2]*, Ricardo Rüther[2] & Carlos R. Rambo[1]

1 *Electrical Materials Laboratory (LAMATE), Department of Electrical Engineering, Universidade Federal de Santa Catarina (UFSC), Florianópolis 88040-970, SC Brazil*
2 *Fotovoltaica UFSC – Solar Energy Research Laboratory, Universidade Federal de Santa Catarina (UFSC), Florianópolis 88056-000, SC Brazil.*

## ABSTRACT

In order to obtain significant increases in the Power Conversion Efficiency (PCE) of solar cells, we argue that future cell research and development should be based on the concomitant improvement of multiple material properties, rather than on the state-of-the-art one- or two-dimensional improvements. In this context, researchers should know, which combined material properties and cell design parameters lead to the highest efficiency increase. For the same objective, it should also be known which relationships in-between these variables have to be adjusted. Such knowledge becomes available by simulation and numerical optimization, which we present for a Perovskite Solar Cell (PSC) in a hypercube space of variables. We prove that its PCE increases principally because of the nonlinearities of this model in the multidimensional space, elucidating the importance of the concomitant variable improvement in the PSC optimization. In the most optimal case, efficiencies of at least 27.6% can be obtained by the simultaneous improvement of the cell's material properties, and light trapping, for a large range of absorber layer thickness of $t_0$ = 160...400 nm. The lower thickness value results in a significant reduction of the cell's lead (Pb) content.

* Corresponding author.
E-mail address: manfredkratzenberg@gmail.com (M. Kratzenberg)

## 1. Introduction

Early research in hybrid solar cells (Huang and Huang, 2014; O'Regan and Grätzel, 1991; Wong et al., 2007) led to the concept of the so-known PSC (Kojima et al., 2009), a hybrid thin-film cell, which absorber layer is constituted by an organic-inorganic compound of semiconductor materials. In searching for new materials and properties of this cell, more recent advances provided the steepest PCE increase in comparison to all other types of cells, reaching the state-of-the-art efficiency of 24.2% for a manufactured prototype (NREL, 2019). Several attempts were accomplished in order to reduce the toxicity inherent to the heavy-metal Pb, present in the absorber layers of high-efficiency PSCs. Such methods consider the complete (Devi and Mehra, 2019; Dixit et al., 2019), or partial substitution of this element (Liu et al., 2018). However, they lead in much lower PCE, if compared to the state-of-the-art efficiency. Our objective is to provide more knowledge and insights about the optimization process of perovskite solar cells and propose high-efficiency designs for future PSC developments. To access this knowledge, and the design concepts, we propose a multidimensional numerical optimization of a cell's drift-diffusion model. Our design concepts show, that the state-of-the-art PCE can be significantly increased, while the Pb content can be reduced up to an order of magnitude.

## 2. Optimization of perovskite solar cells

State-of-the-art PSCs are mostly optimized based on cell prototyping, where the cell's efficiency is increased in small steps as a function of one or two different variables, as permitted by graphical visualization in one- or two-dimensional function spaces. This has the inherent disadvantage that variations in the not measured material properties are unknown. Therefore, an efficiency gain by reason of the improvement of one, or more, of the material properties can be completely or partially nullified by the worsening of other, not controlled, properties. This situation should be better controlled in future cell manufacturing.

The benefits of mathematical modeling of the PSC's efficiency, by use of an analytical model (Kratzenberg et al., 2015; Sun et al., 2015; Taretto et al., 2017), a numerical model (Amu, 2014; Azri et al., 2019; Devi et al., 2018; Dixit et al., 2019; Foster et al., 2014; Iftiquar and Yi, 2016; Kanoun et al., 2019; Liu et al., 2014; Ren et al., 2017; Zhao et al., 2018), or a combination of these models (Agarwal and Nair, 2014, 2015), as used to interpret the cell's drift-diffusion equations, can result in valuable understanding, and accelerate potentially the cell's development process. Similar analytical solutions of the drift-diffusion model were earlier introduced for organic solar cells (Jiang et al., 2015; Chowdhury and Alam, 2014). Finite-difference modeling schemes, which present a better accuracy, because of its higher resolution, are likewise, used to solve these equations (Zandi and Razaghi, 2019), or to simulate the cell's short circuit current density ($J_{sc}$) increase by reason of light trapping (Cai et al., 2015). Some authors also present an efficiency optimization in a one-dimensional optimization of the cells charge conduction layer thickness (Kim and Ohkita, 2017; Zhao et al., 2018). Simulation studies based on the drift-diffusion models lead already in high efficiency values of 25% for single-junction PSCs (Agarwal and Nair, 2014, 2015). The drift-diffusion models simulate the cell's current-voltage curve typically as a function of multiple material properties and the absorber layer thickness. Considering arbitrary material property improvements at different improvement scales, the efficiency of this model can be increased by a multidimensional optimization algorithm. Multidimensional optimizations are state-of-the-art to solve many engineering problems in order to find, e.g. (i) the optimal power flow in electric power grids (Crow, 2009), (ii) the lowest cost design of a wind turbine synchronous generator (Bazzo et al., 2017), or the minimal series resistance in a GaAs solar cell (Algora and Díaz, 2000).

In the analyses of a cell's drift-diffusion model, a multidimensional optimization is useful in order to demonstrate that the concomitant variable improvement results in much higher efficiency increases than in state-of-the-art one- and two-dimensional optimization techniques. Additionally, the multidimensional optimization can lead (i) to a better understanding of the optimization process and (ii) to the proposal of advanced solutions for future high-efficient solar cell design concepts as here proposed.

While the highest state-of-the-art PCE of a manufactured PSC is already 24.2% (NREL, 2019), perovskite solar cells show still a high improvement potential. Considering in a simplified PSC model, only optical recombination losses, the maximal efficiencies of 30% (Yin et al., 2014), 30.88% (Sha et al., 2015) and 31.3% (Martynov et al., 2017) can theoretically be obtained in single-junction PSCs. Furthermore, using zero surface recombination velocities and ideal light trapping in a drift-diffusion model, a PCE of 29.9% can be obtained (Ren et al., 2017). These upper limits are below the highest attainable PCE of single-junction cells, which thermodynamic limit is calculated with 33.7% for the black-body (Shockley and Queisser, 1961), and 33% for the measured irradiation spectrums (Rühle, 2016).

However, in manufactured solar cells, additional recombination losses appear because of a cell's non-ideal material properties. Based on the results from our multidimensional optimizations, we propose cell designs that increase the efficiency of the state-of-the-art manufactured and simulated PSCs. The same proposed cell designs present also a much lower Pb content. The high efficiency of these cell designs is based on high-resolution $J_{sc}$ simulations of efficient light trapping schemes, and we show how the PCE can be increased further, by additional tuning of the cell's material properties. Similar design concepts can be found presently only for manufactured solar cells with much lower PCE, in comparison to state-of-the-art cells (Liu, 2017), or otherwise, only for simulations of the $J_{sc}$ values (Cai et al., 2015). The former author shows that the efficiency of a manufactured PSC, with a 100 nm thick absorber layer, can increase from 5.8% to 7.1% by light trapping; and the latter authors present a finite-difference time-domain simulation (FDTD), which shows that plasmonic light trapping can lead in a 2.15-fold increase of the $J_{sc}$ value, in PSC with a 50 nm thin absorber layer, if compared to a cell of this thickness without light trapping.

## 3. Methodology - Optimization in a hypercube space of variables

To develop a PSC, which shows the highest possible PCE, some fundamental questions should be answered. These are: (i) which multiple material properties and cell design parameters are necessary to be improved concomitantly, (ii) has the value for each of the considered variables to be increased or decreased, and (iii) to which value each of these model variables should be ideally tuned. Using a drift-diffusion model, these questions can be answered by searching the maximal efficiency in a multidimensional function space of material properties and manufacturing parameters.

In the present optimizations, we select an analytical solar cell model, as introduced in (Sun et al., 2015), which efficiency we maximized by a numerical optimization method and the simplifications of this model are discussed in Appendix B1.1. The selected model and method reduce significantly the computational optimization cost if compared to a numerical PSC model and random search methods. This enabled us to accomplish a large set of optimizations, using different constraints in each of these optimizations, for the model's variable set. The approximations made by the selected analytical model results in a low fitting deviation of only 0.1%, between the measured and the simulated PCE values (Sun et al., 2015). This small deviation is insignificant considering that outdoor spectral variations can lead to PCE uncertainties of 3% in thin-film cells (Rüther et al., 2002). The used model was elaborated in from (i) the second-order Poisson differential equation and its two boundary equations and (ii) a further set of four second-order drift-diffusion differential equations, and its boundary equations, which describe the cell's electron and hole transports in the dark and under exposition to light (Sun et al., 2015).

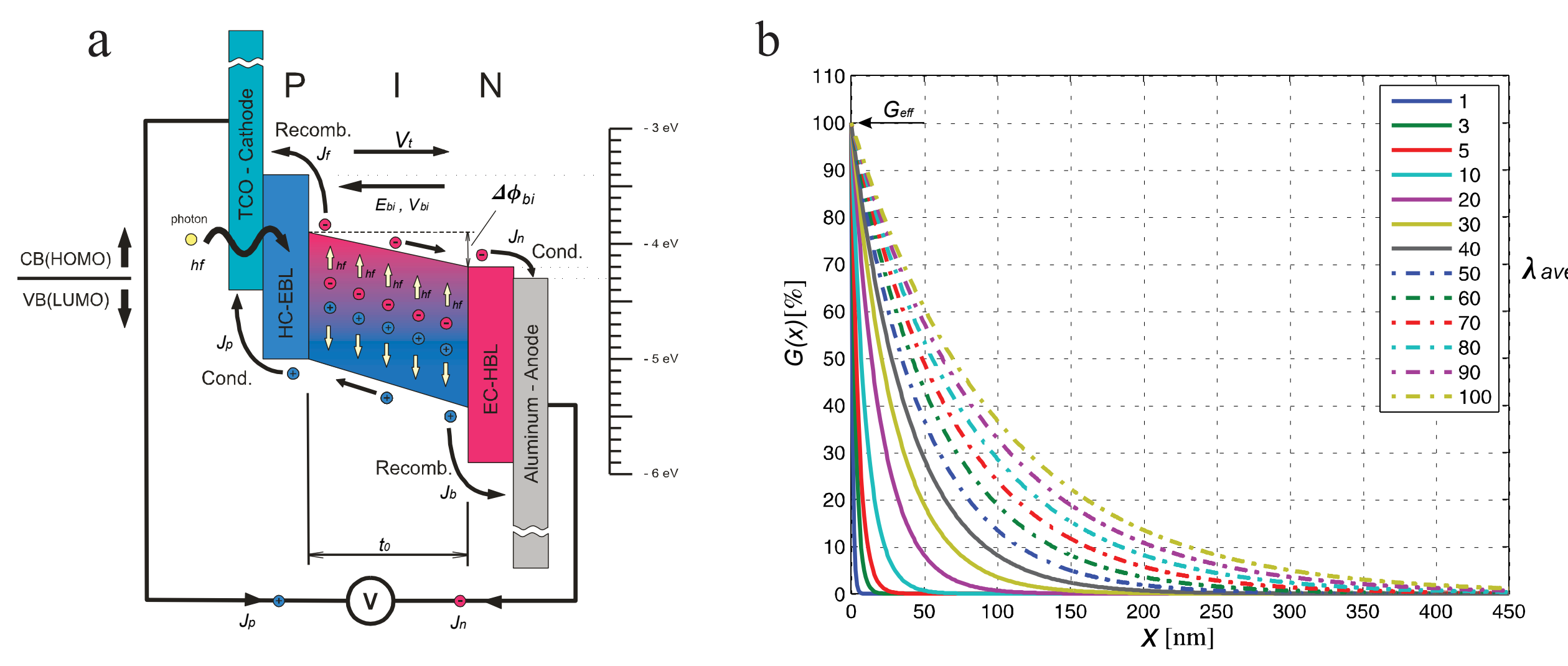

**Fig. 1.** Energy band diagram and normalized charge generation: (a) Energy potential diagram and charge carrier trajectories of a planar pin-type heterojunction perovskite solar cell, as modeled in (Sun et al. 2015), with the following components from left to right: (i) transparent top cover (here not visualized); (ii) cathode layer made of a Transparent Conductive Oxide (TCO); (iii) p-type hole conduction and electron blocking layer, made of organic material PEDOT:PSS; (iv) i-type intrinsic charge generation layer with thickness $t_0$ = 450 nm made of the hybrid perovskite material; (v) n-type electron conduction and hole blocking layer, made of organic material PCBM; (vi) anode layer made of aluminum; adapted from supplementary information in (Sun et al. 2015). (b) Approximation of the normalized charge generation profiles $G(x)$ [$cm^{-2}s^{-1}$] per solar cell area and time as a function of the penetration depth ($x$ = 0…450 nm) on the abscissa, calculated for several of the considered average optical decay lengths of $\lambda_{ave}$ = 5 to 150 nm.

A detailed charge transport scheme, of the used pin-type PSC, is presented in Fig. 1(a). Based on FDTD simulations (Cai et al., 2015), our optimized solar cell architectures present light trapping nanoparticles in its thin absorber layers, which increase its short circuit current density ($J_{sc}$), if compared to the $J_{sc}$ = 23 mA/cm$^2$ without light trapping, in a large range of absorber layer thicknesses.

3.1 Modeling of the optimization

In this work, we propose the numerical optimization of the efficiency of PSC's analytical model in a multidimensional function space, which is specified by the following optimization problem

$$\eta_{max} \rightarrow \max [\ \eta_i (X_{1,i} \ldots X_{9,i})\ ]\ \ ;\ \ i = 1 \ldots N\ , \tag{1}$$

where the efficiency $\eta_i$ in the *i*-th maximization iteration is a function of the nine PSC model variables ($X_{1,i}...X_{9,i}$), which built-up a nine-dimensional hypercube space, and where the maximum efficiency ($\eta_{max}$) is obtained after $i = 1...N$ model simulations. In a single iteration of this optimization, each of the nine model variables is set up to a different value, and therefore, each iteration results in a simulation ($i$), where the variable values are selected by the optimization algorithm. As initial model variables ($X_1...X_9$) we consider the values as obtained by the optimization in (Sun et al., 2015). In each of the $N$ steps, the Matlab™ optimization algorithm *fmincon* improves the values of the whole set of model variables by a Nonlinear Programming (NP) optimization technique (Byrd et al., 1999), which is based on an interior point (IPM) and line search optimization methods (LSM), in order to obtain the maximal possible model efficiency. When the cell's efficiency increase is below a specified threshold value, the optimization algorithm considers that the maximal cell efficiency is obtained, and it stops the optimization process (Fig. A.2). For each of the $i = 1 \ldots N$ model simulations, a new calculation of (i) the J-V curves, (ii) the maximum power point (MPP) power, and (iii) the cell's efficiency ($\eta_i$) are accomplished, using equation (2). This PCE is calculated as a function of nine model variables $X_{1,i}$ to $X_{9,i}$, which are constituted by eight quantum physical material properties and the absorber layer thickness.

$$\eta_{MPP,i} = (\ U_{MPP,i}\ J_{MPP,i}(X_{1,i}...X_{9,i})\ )\ /\ G_{AM1.5} = P_{MPP,i}\ /\ G_{AM1.5} \tag{2}$$

In equation (2) $G_{AM1.5}$ = 100 mW/cm² is the normalized solar irradiation at Air Mass 1.5. The PSC's maximum output power density $P_{MPP,i}$ [mW/m$^2$] of the $i$-th optimization step is obtained by the maximization of its power curve in $k = 1 \ldots M$ iterative steps as follows

$$P_{MPP,i} \rightarrow \max \left(P_k \left( J_{light,1} \left(G_{AM1.5}, V_k, X_{1,i} \ldots X_{9,i}\right) \ldots J_{light,M} \left(G_{AM1.5}, V_k\right) \right) \;;\; k = 1 \ldots M \right., \quad (3)$$

where each $P_{MPP,i}$ is related to PSC's material properties in the $i$-th optimization step. In order to calculate $P_{MPP,i}$ the cell's analytical model uses the values of the whole set of model variables ($X_{1,i} \ldots X_{9,i}$) as well as the terminal voltage ($V_k$) and the cell's temperature as input variables to calculate its current density $J_{light,k}$ under exposure to reference light ($G_{AM1.5}$). The material properties are not modified in this second optimization. We remark that the power optimization of the equation (3) is nested in the efficiency optimization (equation 1). This nested optimization process is subject to the following specific boundary conditions

$$X_{j,min} \leq X_j \leq X_{j,max} \quad ; \quad j = 1 \ldots 9\,, \quad (4)$$

where $X_{j,max}$ and $X_{j,min}$ are the maximal and minimal constraints for each one of the nine model variables $X_j$ to be optimized. Considering a large number of optimizations, each under different boundary conditions, specific constraints of the model variables in an arbitrary optimization process are given by equation (5). Each single optimization process considers a variable specific range ($X_{j-min}$ ...$X_{j,max}$), in which it is allowed for the optimization algorithm to modify each of the $j$ = 1…9 model variables. Here we only consider for the built-in voltage ($V_{bi}$), the absorber layer thickness ($t_0$), and the average optical decay length ($\lambda_{ave}$), variable specific constrains, as defined by the equations (6), (7) and (9). As we cannot know which property improvements will be possible in the future cell development, we set up arbitrary defined and adjustable constraints for remaining material properties, by a proposed variable boundary expansion factor ($f_B$), as given by the following equation

$$X_{j-min} = (1/f_B)\, X_{j,me} \leq X_j \leq X_{j,me}\, (f_B) = X_{j,max} \quad ; \quad j = 1 \ldots 9\,, \quad (5)$$

where the terms $f_B$ and $1/f_B$ specify individual amplification and reduction factors in a single optimization process. The application of $f_B$ results in the values of the upper and lower boundary limits of the model variables ($X_{j\text{-}min}$ and $X_{jmin}$) in a specific efficiency optimization. In equation (5) $X_{j,me}$ represents the not expanded setup configuration of the model variables, as measured in (Sun et al., 2015), and $X_{j\text{-}min}$ and $X_{j\text{-}max}$ are the minima and the maxima limits as specified in equation (4). Built-in voltages with values ($V_{bi}$ > 1.4 V) do not lead to any further increase of the cell's efficiency (Fig. 2(b)). Hence, the inequality constraint equation (6) is imposed additionally to the optimization in order to avoid unrealistic high values of this variable. In fact, $f_B$ is used in this case to increase in small steps $V_{bi}$ to the maximal value of $X_{max}$ = 1.4 V (Tables 1 to 3).

$$0 \leq V_{bi} \leq 1.4 \text{ V} \tag{6}$$

Furthermore, we consider that each of the different available coating techniques of the absorber layer needs a specific minimal thickness ($t_{0\text{-}min}$) in order to avoid losses related to pinholes (Qiu et al., 2016) and conduction effects (Atwater and Polman, 2010). Therefore, we set up an additional constraint for $t_0$, which considers (i) an adjustable minimum thickness ($t_{0\text{-}min}$) as the lower boundary in an optimization process, while (ii) the upper limit of $t_0$ is here assumed to be 1 µm as specified by the inequality constraint equation (7).

$$t_{0\text{-}min} \geq t_0 \geq 1 \mu\text{m} \tag{7}$$

In the Beer-Lambert law $\lambda_{ave}$ stands in a defined relationship with $t_0$, for the consideration of low reflection losses of approximately 1% (Fig. 1b). This relation is expressed by equation (8) for the case without light trapping (Sun et al., 2015).

$$m = t_0 / \lambda_{ave} \tag{8}$$

With the setup values for $t_0$ and $\lambda_{ave}$ (section 3.2) the value of $m$ = 4.5, is obtained, which should also be considered for cells with light trapping, considering similar losses. Light trapping reduces the average optical decay length ($\lambda_{ave}$) and improves at the same time the $J_{sc}$ (Cai et al., 2015). We use the calculated $m$ = 4.5 in order to specify an additional equality constraint (equation 9), here defined as the m-constrained determination of $\lambda_{ave}$. In this context we calculate $\lambda_{ave}(t_0)$ for light trapping as a function of $t_0$, solving equation (8) for $\lambda_{ave}$ as follows

$$\lambda_{ave}(t_0) = t_0 / m = t_0 / 4.5 \,. \tag{9}$$

If $\lambda_{ave}$ is instead in the same form constrained as the further model variables, by use of the $f_B$ factor, a relationship appears, as defined by the vertex lines in Figs. 2e and 2f, and Table A.3.2, which results in lower $\lambda_{ave}$ but higher optimized efficiency values (Tables 1 to 3 – last two columns). However, improved absorption properties, and therefore, a reduced $\lambda_{ave}$, as obtained by light trapping, should stand in a similar relationship to the absorber layer thickness as in a cell without light trapping, by reason of the Beer-Lambert law (Fig. 1b). Therefore, we calculate an $m$ - constrained $\lambda_{ave}$ in equation (9), which results in a lower PCE, and presents, therefore, a more conservative formulation. In our $m$ - constraint optimizations the remaining material properties are constrained by $f_B$ - factor and equations (6) and (7). Whereas higher efficient light trapping schemes by shape-optimized plasmonic nanoparticles are possible (Kakavelakis et al., 2017), the best light-trapping scheme, as here adopted, consider spherical silver (Ag) nanoparticles, which are localized at the backside of the absorber layer (Cai et al., 2015). The short circuit current density for this cell configuration is, for a large thickness range, higher than the value as obtained in (Sun et al., 2015). Therefore, a short circuit current density correction factor for light trapping should be considered as a function of the absorber layer thicknesses, which is calculated as follows

$$\mathrm{CF}_{lt}(t_0) = J_{sc\text{-}Cai}(t_0) \,/\, J_{sc\text{-}Sun} \,. \tag{10}$$

As without light trapping both authors presented a similar short circuit current density for the same absorber material, and an absorber layer thickness of 400 nm, we can directly adopt short circuit current densities as obtained by FDTD simulations in (Cai et al., 2015), in our simulation setup.

### 3.2 Setup conditions

1 We used the solar cell properties and $t_0$ as obtained in (Sun et al., 2015) as principal setup conditions.
2 The authors calculated by use of the OTM method, the fundamental optical parameters, which are: (i)
3 the effective generation of charge carriers, $G_{eff}$ = 1.4356 x $10^{13}$ $cm^{-3}s^{-1}$, and (ii) the average optical
4 decay length $\lambda_{ave}$ = 100 nm, which are both specific material properties that are independent of absorber
5 layer thickness. By the derived equation for the total generation of free charge carriers (equations A.1
6 and A.2) a $G_{max}$ of 1.4356 x $10^{17}$ $cm^{-2}s^{-1}$ is calculated, which coincides with the measured short circuit
7 current density of $qG_{max}$ = 23 mA/cm², where $q$ is the electric charge energy. Built-in voltage and the
8 diffusion coefficients are $V_{bi}$ = 0.78 V, and $D_n = D_p$ = 0.05 $cm^2$/s. The authors obtained an ideal $t_0$ of
9 450 nm using a one-dimensional optimization of this variable. After the manufacturing of the PSC
10 with the optimized absorber layer thickness, the authors obtained by a curve fitting: (i) the surface
11 recombination velocity of electrons in the front charge transport layer $s_f = s_n$ = 200 cm/s, (ii) the surface
12 recombination velocity of holes in the back charge transport layer $s_p = s_b$ = 19.2 cm/s, and (iii) the
13 number of the excess minority carrier concentrations of electrons and holes ($\Delta n$ = 8.426 x $10^6$ $cm^{-3}$, $\Delta p$
14 = 1.3003 x $10^8$ $cm^{-3}$). Furthermore, they obtained an optimized PCE of 15.7%, in a cell with an open-
15 circuit voltage of $V_{0C}$ = 0.87 V (Table 4 - design 6). The measured semiconductor temperature was
16 300.56 K, which results in a thermal voltage of $V_t = k_B T/q$ = 25.9 mV. As short circuit current densities
17 $qG_{max}$ we adopted the values as obtained by the FDTD simulations in (Cai et al., 2015) for light
18 trapping with spherical nanoparticles, as a function of the absorber layer thickness of $t_0$ = 50…400 nm
19 (equation 10).
20
21 **4 Results**
22 We propose (i) a multidimensional optimization of a solar cell's efficiency, as based on its drift-
23 diffusion model, and present additionally, (ii) a complete set of the accomplishable one- and two-
24 dimensional sensitivity analyses, as obtained by the possible variable combinations. Based on our
25 results, we propose different cell designs, which increase the efficiency of state-of-the-art simulated

1 and manufactured cells. Figs. 2(a) to 2(e) show some typical one- and two-dimensional, sensitivity

2 analyses, where the efficiency is calculated as a function of the improvement of its model variables.

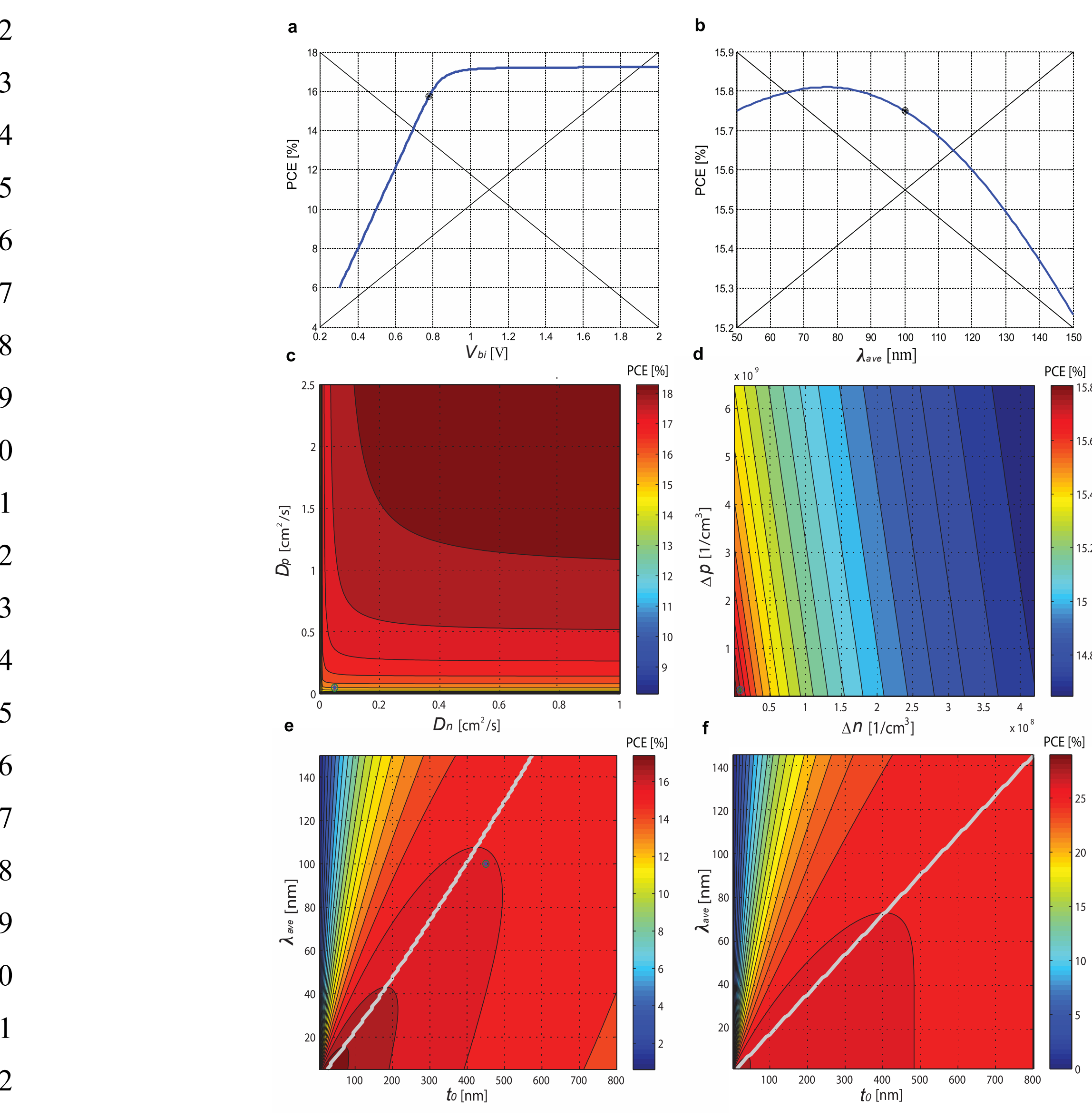

**Fig. 2.** One- and two-dimensional sensitivity analyses, showing the model efficiency as a function of: (a) the average optical decay length $\lambda_{ave}$; (b) the built-in voltage $V_{bi}$; (c) the diffusion coefficients of electrons and holes $D_n$ and $D_p$; (d) the excess concentration of electrons and holes $\Delta n$ and $\Delta p$; and finally, in (e) and (f), the absorber layer thickness $t_0$, and the average optical decay length, where the grey vertex line shows the maximal attainable efficiency values in these two-dimensional presentations; (a) – (e) the remaining model variables are set up to the values as obtained in (Sun et al. 2015); (f) the values of the remaining model variables are set up to the values as obtained in our $f_B$ constrained multidimensional optimization for $f_B$ = 160; the measured efficiency values of 15.7% in (a) – (e) appear as a star within a surrounded circle.

For the visualization of the efficiency gains, as obtained by the possible combinations of nine model variables, a large set of 45 such figures would be necessary. However, these figures can be substituted by Table A.1, which shows only the highest attainable PCE values of those analyses, as a function of different boundary expansion factors $f_B$ (equation 5). For the highest variable improvement factor ($f_B$ = 160) the best one- and two-dimensional efficiency optimizations increase the PCE from 15.7% to 18.1% and 20%. The Figs. 2(e) and 2(f) compare two similar two-dimensional sensitivity analyses using the same model variables. While in Fig. 2(e) the remaining variables confer to our setup condition, in Fig. 2(f) these variables are configured with the ideal values as obtained by a multidimensional optimization with $f_B$ = 160. Both figures consider light trapping with independently adjustable $t_0$ and $\lambda_{ave}$, but the multidimensional optimization results in higher efficiency values. It can be observed from the slight shift of the grey vertex lines in these figures that for arbitrary values of $t_0$, the multidimensional optimization demands an improved light trapping, by reason of the lower optimal values of $\lambda_{ave}$ for the same $t_0$, if compared to the two-dimensional optimization. Based on the results of our $f_B$ - constraint multidimensional optimization, as presented in Table A.2, we conclude that a variable relationship that results in maximal efficiency vertex lines does only appear in-between the variables $t_0$ and $\lambda_{ave}$, and not in-between the further model variables.

Fig. 3 presents six sets of 160 multidimensional optimizations ($f_B$ = 1…160), for six different absorber layer thicknesses. For these *m*-constrained optimizations the variables $V_{bi}$, $t_0$, $\lambda_{ave}$ and the short circuit current densities are set up by the equations (6), (7), (9) and (10), while the remaining variables are $f_B$ - constrained. As specified in equation (7), we consider that each coating technique of the absorber has its own inherent minimum thickness ($t_{0\text{-}min}$) for the deposition of perfect absorber layers (Appendix section A.4), and therefore, we present optimized designs for different absorber layer thicknesses. Light trapping with spherical nanoparticles (solid curves) presents a significant efficiency advantage in comparison to the PSC without light trapping (dotted curve) in almost all development states that are expressed by the $f_B$ factor. In a 160 nm thick absorber layer, such a light-trapping shows only 0.1...0.2% lower efficiencies, in comparison to the configuration with the highest efficiencies ($t_0$ = 400 nm). Additionally, a PSC with $t_0$ = 160 has the advantage of presenting 2.5-fold lower Pb content, and therefore, we give special emphasis to this cell design in our discussions.

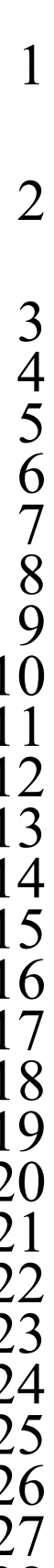

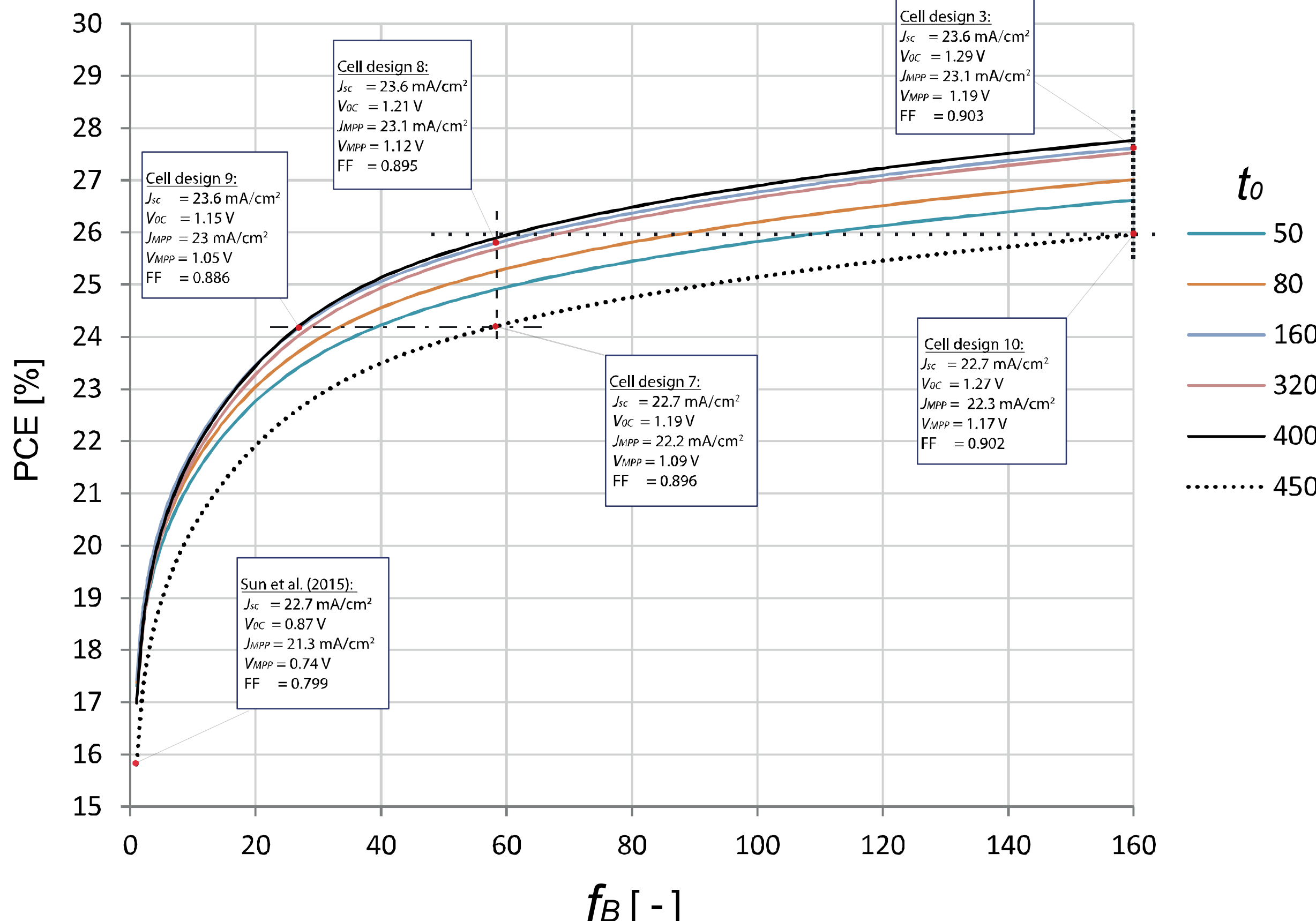

**Fig. 3.** Optimized efficiency as a function of the 160 boundary expansion factors ($f_B$ = 1…160, ∊ ℕ) for six absorber layer thicknesses in the range of $t_0$ = 50…450 nm, optimization variables ($D_n$, $D_p$, $V_{bi}$, $s_f$, $s_b$), while $\lambda_{ave}(t_0) = t_0/4.5$ is the calculated average optical decay length: (i) without light trapping, $t_{0\text{-}min}$ = 450 nm, $\lambda_{ave}$ = 100 nm; (ii) light trapping with spherical plasmonic nanoparticles, $t_0$ = 50….400 nm, with corrected short circuit current densities. Used $qG_{max}$ values, in unit [mA/cm$^2$], as obtained in (Cai et al. 2015 and Sun et al. 2015): 22.5 ($t_0$ = 50 nm), 23.05 (80 nm), 23.9 (160 nm), 24.18 (320 nm), 24.5 (400 nm), 23 (450 nm, without light trapping).

In Tables 1, 2 and 3, we show an alternative representation for some optimizations in Fig. 3, which also show the values of the optimized material properties. While Table 1 considers the state-of-the-art absorber layer thickness of 450 nm and no light trapping, Tables 2 and 3 present the results for thinner absorber layers of 400 nm and 160 nm, considering a $J_{sc}$ as obtained by light trapping with spherical nanoparticles and a corrected $\lambda_{ave}$ by equation (9). For the interested reader, we additionally show the PCE and $\lambda_{ave}$ values, as obtained in optimizations with $f_B$ - constrained $\lambda_{ave}$ setups (equation 5). However, these values should be observed with caution, because of the related low reflection losses, which might be unrealistic.

Figs. 4a and 4b present, the J-V curves, and the power curves, of nine different high-efficiency design proposals. The figure visualizes in more detail, some of the simulated design proposals as presented in Fig. 3. Highest efficiency designs (1) to (5) show light trapping for $t_0$ = 50…400 nm, with $f_B$ = 160. While design (6) presents the cell as proposed is (Sun et al., 2015), (7) confers, as an example, to the cell with the state-of-the-art efficiency, considering improved material properties and no light trapping. Design (8) shows how (7) can be improved by light trapping, and (9) shows a similar efficiency as (7), with less ideal properties, but with light trapping. Table 4 presents the J-V performance parameters of the design proposals as shown in Figs. 3 and 4, and Table 5 visualizes the relative improvements, considering in each row the comparison of two of the design proposals, as presented in Table 4.

**Table 1**

Optimized model variables (boldface symbols) as obtained from optimizations in a seven-dimensional function space considering eleven optimizations with specific boundary expansion factors $f_B$, obtaining an ideal PSC of 25.96%, for a cell with a fixed $t_0$ of 450 nm, which do not use light trapping ($\lambda_{ave}$ = 100 nm); $qG_{max}$ = 23 mA/cm$^2$.

| $f_B$ [-] | **$s_f$** [cm/s] | **$s_b$** [cm/s] | **$\Delta n$** [1/cm$^3$] | **$\Delta p$** [1/cm$^3$] | **$V_{bi}$** [V] | **$D_n$** [cm²/s] | **$D_p$** [cm²/s] | $\mu_n$ [cm²/Vs] | $\mu_p$ [cm²/Vs] | $t_0$ [nm] | **$\lambda_{ave}$** [nm] | $\eta$ [%] | **$\lambda_{ave}^*$** [nm] | $\eta^*$ [%] |
|---|---|---|---|---|---|---|---|---|---|---|---|---|---|---|
| *5.00* | 40.00 | 3.84 | 8.43E+06 | 1.30E+08 | 0.80 | 0.25 | 0.25 | 9.65 | 9.65 | *450.00* | 100.00 | *18.96* | 48.13 | 19.15 |
| *10.00* | 20.00 | 1.92 | 8.43E+06 | 1.30E+08 | 0.82 | 0.50 | 0.50 | 19.30 | 19.30 | *450.00* | 100.00 | *20.34* | 10.00 | 20.57 |
| *20.00* | 10.00 | 0.96 | 8.43E+06 | 1.30E+08 | 0.86 | 1.00 | 1.00 | 38.61 | 38.61 | *450.00* | 100.00 | *21.92* | 5.00 | 22.17 |
| *30.00* | 6.67 | 0.64 | 8.43E+06 | 1.30E+08 | 0.90 | 1.50 | 1.50 | 57.91 | 57.91 | *450.00* | 100.00 | *22.88* | 3.33 | 23.14 |
| *40.00* | 5.00 | 0.48 | 8.43E+06 | 1.30E+08 | 0.94 | 2.00 | 2.00 | 77.22 | 77.22 | *450.00* | 100.00 | *23.49* | 2.50 | 23.76 |
| *60.00* | 3.33 | 0.32 | 8.43E+06 | 1.30E+08 | 1.01 | 3.00 | 3.00 | 115.82 | 115.82 | *450.00* | 100.00 | *24.26* | 1.67 | 24.53 |
| *80.00* | 2.50 | 0.24 | 8.43E+06 | 1.30E+08 | 1.09 | 4.00 | 4.00 | 154.43 | 154.43 | *450.00* | 100.00 | *24.76* | 27.89 | 25.04 |
| *100.00* | 2.00 | 0.19 | 8.43E+06 | 1.30E+08 | 1.17 | 5.00 | 5.00 | 193.04 | 193.04 | *450.00* | 100.00 | *25.14* | 26.85 | 25.43 |
| *120.00* | 1.67 | 0.16 | 8.43E+06 | 1.30E+08 | 1.25 | 6.00 | 6.00 | 231.65 | 231.65 | *450.00* | 100.00 | *25.46* | 25.85 | 25.75 |
| *140.00* | 1.43 | 0.14 | 8.43E+06 | 1.30E+08 | 1.32 | 7.00 | 7.00 | 270.26 | 270.26 | *450.00* | 100.00 | *25.73* | 25.01 | 26.02 |
| *160.00* | 1.25 | 0.12 | 8.43E+06 | 1.30E+08 | 1.40 | 8.00 | 8.00 | 308.87 | 308.87 | *450.00* | 100.00 | *25.96* | 0.63 | 26.26 |

Observations: The values in italic formatted numbers correspond to the values of the lower curve in Fig. 3; * values obtained in $f_B$ – constrained determination of $\lambda_{ave}$

**Table 2**

Optimized model variables (boldface symbols) and efficiency in seven-dimensional optimizations as a function of several boundary extension factors $f_B$ for a PSC with $t_0$ = 400 nm, where light trapping by use of spherical nanoparticles compensates the reduced absorption in this absorber layer; $qG_{max}$ = 24.5 mA/cm$^2$.

| $f_B$ [-] | **$s_f$** [cm/s] | **$s_b$** [cm/s] | **$\Delta n$** [1/cm$^3$] | **$\Delta p$** [1/cm$^3$] | **$V_{bi}$** [V] | **$D_n$** [cm²/s] | **$D_p$** [cm²/s] | $\mu_n$ [cm²/Vs] | $\mu_p$ [cm²/Vs] | $t_0$ [nm] | **$\lambda_{ave}$** [nm] | $\eta$ [%] | **$\lambda_{ave}^*$** [nm] | $\eta^*$ [%] |
|---|---|---|---|---|---|---|---|---|---|---|---|---|---|---|
| *5.00* | 40.00 | 3.84 | 8.43E+06 | 1.30E+08 | 0.80 | 0.25 | 0.25 | 9.65 | 9.65 | *400.00* | 88.89 | *20.29* | 41.76 | 20.50 |
| *10.00* | 20.00 | 1.92 | 8.43E+06 | 1.30E+08 | 0.82 | 0.50 | 0.50 | 19.30 | 19.30 | *400.00* | 88.89 | *21.76* | 8.89 | 22.01 |
| *20.00* | 10.00 | 0.96 | 8.43E+06 | 1.30E+08 | 0.86 | 1.00 | 1.00 | 38.61 | 38.61 | *400.00* | 88.89 | *23.45* | 4.44 | 23.72 |
| *30.00* | 6.67 | 0.64 | 8.43E+06 | 1.30E+08 | 0.90 | 1.50 | 1.50 | 57.91 | 57.91 | *400.00* | 88.89 | *24.47* | 2.96 | 24.76 |
| *40.00* | 5.00 | 0.48 | 8.43E+06 | 1.30E+08 | 0.94 | 2.00 | 2.00 | 77.22 | 77.22 | *400.00* | 88.89 | *25.13* | 2.22 | 25.42 |
| *60.00* | 3.33 | 0.32 | 8.43E+06 | 1.30E+08 | 1.01 | 3.00 | 3.00 | 115.82 | 115.82 | *400.00* | 88.89 | *25.95* | 1.48 | 26.25 |
| *80.00* | 2.50 | 0.24 | 8.43E+06 | 1.30E+08 | 1.09 | 4.00 | 4.00 | 154.43 | 154.43 | *400.00* | 88.89 | *26.48* | 24.56 | 26.79 |
| *100.00* | 2.00 | 0.19 | 8.43E+06 | 1.30E+08 | 1.17 | 5.00 | 5.00 | 193.04 | 193.04 | *400.00* | 88.89 | *26.90* | 23.71 | 27.17 |
| *120.00* | 1.67 | 0.16 | 8.43E+06 | 1.30E+08 | 1.25 | 6.00 | 6.00 | 231.65 | 231.65 | *400.00* | 88.89 | *27.23* | 22.83 | 27.54 |
| *140.00* | 1.43 | 0.14 | 8.43E+06 | 1.30E+08 | 1.32 | 7.00 | 7.00 | 270.26 | 270.26 | *400.00* | 88.89 | *27.52* | 22.15 | 27.83 |
| *160.00* | 1.25 | 0.12 | 8.43E+06 | 1.30E+08 | 1.40 | 8.00 | 8.00 | 308.87 | 308.87 | *400.00* | 88.89 | *27.76* | 21.50 | 28.08 |

Observations: The values in italic formatted numbers correspond to the cell with $t_0$ = 400 nm in Fig. 3; * values obtained in $f_B$ – constrained determination of $\lambda_{ave}$

**Table 3**
Optimized model variables (boldface symbols) and efficiencies in seven-dimensional optimizations as a function of several boundary extension factors $f_B$ for a PSC with $t_0$ = 160 nm, where light trapping by spherical nanoparticles compensates the reduced absorption in this absorber layer; $qG_{max}$ = 23.9 mA/cm$^2$.

| $f_B$ [-] | $\mathbf{s_f}$ [cm/s] | $\mathbf{s_b}$ [cm/s] | $\mathbf{\Delta n}$ [1/cm$^3$] | $\mathbf{\Delta p}$ [1/cm $^3$] | $\mathbf{V_{bi}}$ [V] | $\mathbf{D_n}$ [cm²/s] | $\mathbf{D_p}$ [cm²/s] | $\mu_n$ [cm²/Vs] | $\mu_p$ [cm²/Vs] | $t_0$ [nm] | $\boldsymbol{\lambda_{ave}}$ [nm] | $\eta$ [%] | $\boldsymbol{\lambda_{ave}}^*$ [nm] | $\eta^*$ [%] |
|---|---|---|---|---|---|---|---|---|---|---|---|---|---|---|
| *5.00* | 40.00 | 3.84 | 8.43E+06 | 1.30E+08 | 0.80 | 0.25 | 0.25 | 9.65 | 9.65 | *160.00* | 35.56 | *20.46* | 7.11 | 21.20 |
| *10.00* | 20.00 | 1.92 | 8.43E+06 | 1.30E+08 | 0.82 | 0.50 | 0.50 | 19.30 | 19.30 | *160.00* | 35.56 | *21.85* | 3.56 | 22.67 |
| *20.00* | 10.00 | 0.96 | 8.43E+06 | 1.30E+08 | 0.86 | 1.00 | 1.00 | 38.61 | 38.61 | *160.00* | 35.56 | *23.46* | 1.78 | 24.34 |
| *30.00* | 6.67 | 0.64 | 8.43E+06 | 1.30E+08 | 0.90 | 1.50 | 1.50 | 57.91 | 57.91 | *160.00* | 35.56 | *24.43* | 1.19 | 25.35 |
| *40.00* | 5.00 | 0.48 | 8.43E+06 | 1.30E+08 | 0.94 | 2.00 | 2.00 | 77.22 | 77.22 | *160.00* | 35.56 | *25.06* | 0.89 | 26.00 |
| *60.00* | 3.33 | 0.32 | 8.43E+06 | 1.30E+08 | 1.01 | 3.00 | 3.00 | 115.82 | 115.82 | *160.00* | 35.56 | *25.84* | 0.59 | 26.82 |
| *80.00* | 2.50 | 0.24 | 8.43E+06 | 1.30E+08 | 1.09 | 4.00 | 4.00 | 154.43 | 154.43 | *160.00* | 35.56 | *26.37* | 9.08 | 27.36 |
| *100.00* | 2.00 | 0.19 | 8.43E+06 | 1.30E+08 | 1.17 | 5.00 | 5.00 | 193.04 | 193.04 | *160.00* | 35.56 | *26.77* | 0.60 | 27.77 |
| *120.00* | 1.67 | 0.16 | 8.43E+06 | 1.30E+08 | 1.25 | 6.00 | 6.00 | 231.65 | 231.65 | *160.00* | 35.56 | *27.10* | 0.90 | 28.11 |
| *140.00* | 1.43 | 0.14 | 8.43E+06 | 1.30E+08 | 1.32 | 7.00 | 7.00 | 270.26 | 270.26 | *160.00* | 35.56 | *27.37* | 8.44 | 28.40 |
| ***160.00*** | 1.25 | 0.12 | 8.43E+06 | 1.30E+08 | 1.40 | 8.00 | 8.00 | 308.87 | 308.87 | ***160.00*** | 35.56 | ***27.61*** | 0.71 | 28.65 |

Observations: (i) The values in boldface formatted numbers correspond to cell design 3 in Fig.4, and high-efficiency cell in Fig. 5; (ii) the values in italic formatted numbers correspond to the cell with $t_0$ = 160 nm in Fig. 3; $^*$values obtained in $f_B$ – constrained determination of $\lambda_{ave}$.

3 The curves in Fig. 3 visualize how the efficiency of a PSC in a certain development state can be
4 optimized by (i) the improvement of the material properties, as expressed by the $f_B$ factor; (ii) light
5 trapping for a constant $f_B$ factor, as used for $t_0$ < 450 nm; and (iii) the combination of these methods.
6 While material property improvements result in efficiency growths almost solely based on the
7 increase of $V_{MPP}$; light trapping increases the efficiency as a function of both, a $V_{MPP}$ rise, and
8 predominantly, an increase of $J_{MPP}$. Such improvements can be discussed comparing the optimization
9 results in Table 4 (Fig.4), which results in Table 5. Considering that the cell with the state-of-the-art
10 efficiency of 24.2 % is based on an effective light trapping scheme (Figure 3, design 9), a relative
11 efficiency rise of 14.6 % (Table 5 - line 13) can be obtained as a function of further material property
12 improvements. This increase is almost solely based on a 14.1 % rise of $V_{MPP}$ and results in the
13 proposed cell design 3.

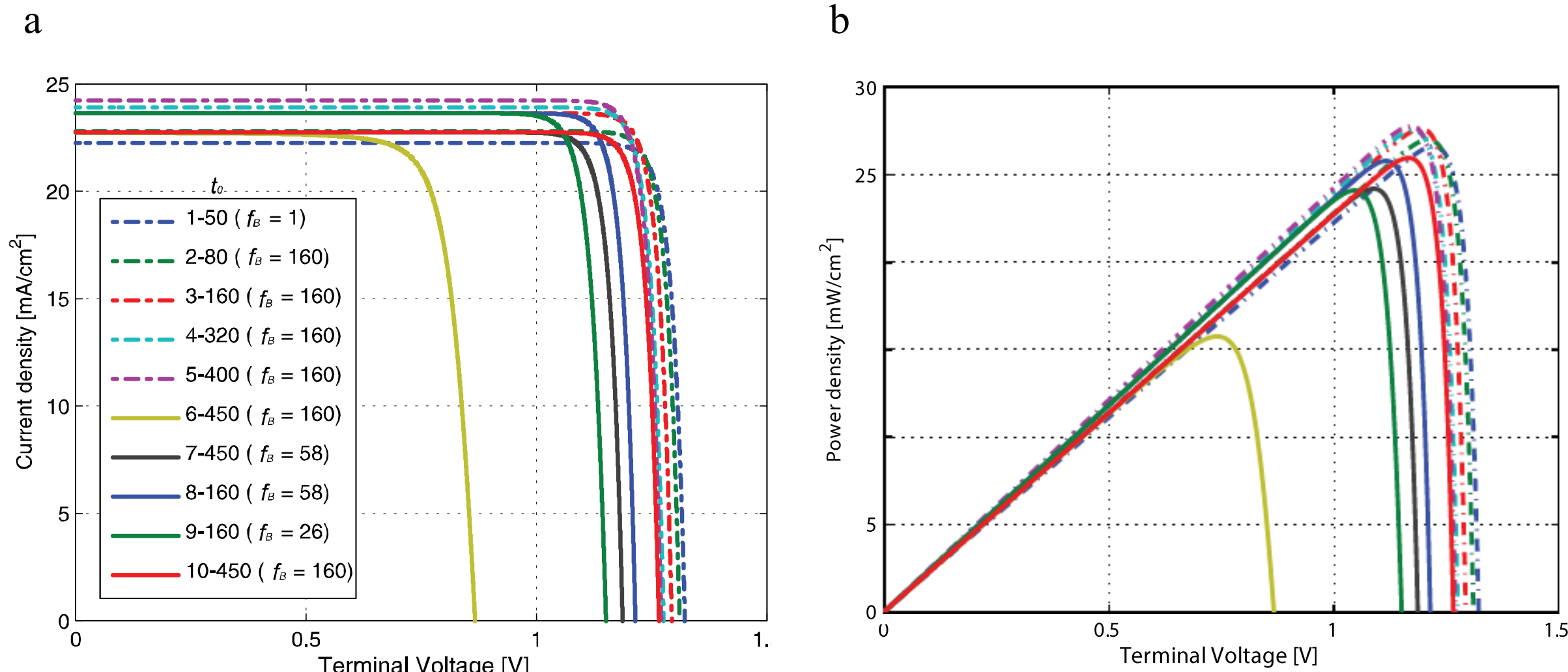

**Fig. 4.** (a) J-V and (b) power curves of different solar cell designs, as obtained by the multidimensional optimizations, as specified by the absorber layer thickness ($t_0$) and boundary amplification factor ($f_B$): (1) $t_0$ = 50 nm, $f_B$ = 160; (2) $t_0$ = 80 nm, $f_B$ = 160; (3) $t_0$ = 160 nm, $f_B$ = 160; (4) $t_0$ = 320 nm, $f_B$ = 160; (5) $t_0$ = 400 nm, $f_B$ = 160; (6) $t_0$ = 450 nm, $f_B$ = 1 (Sun et al., 2015); (7) $t_0$ = 450 nm, $f_B$ = 58 ≈ state-of-the-art efficiency value; (8) $t_0$ = 160 nm, $f_B$ = 58; (9) $t_0$ = 160 nm, $f_B$ = 26 ≈ state-of-the-art efficiency value, (10) $t_0$ = 450, $f_B$ = 160.

**Table 4**

Optimization setup configurations and perovskite performance parameters extracted from the J-V curve characteristics for the cell designs as presented in Figs. 3 and 4.

| **Cell design** | $f_B$ [ - ] | $t_0$ [nm] | $\lambda_{ave}$ [nm] | $q\ G_{max}$ (#) [mA/cm$^2$] | $J_{SC}$ [mA/cm$^2$] | $V_{0C}$ [V] | $J_{MPP}$ [mA/cm$^2$] | $V_{MPP}$ [V] | FF [ - ] | PCE [%] |
|---|---|---|---|---|---|---|---|---|---|---|
| **1** | 160 | 50 | 11.1 | 22.5 | 22.3 | 1.32 | 21.8 | 1.22 | 0.905 | 26.6 |
| **2** | 160 | 80 | 17.8 | 23.1 | 22.8 | 1.31 | 22.3 | 1.21 | 0.904 | 27.0 |
| **3** | 160 | 160 | 35.6 | 23.9 | 23.6 | 1.29 | 23.1 | 1.19 | 0.903 | 27.6 |
| **4** | 160 | 320 | 71.1 | 24.2 | 23.9 | 1.28 | 23.4 | 1.18 | 0.902 | 27.5 |
| **5** | 160 | 400 | 88.9 | 24.5 | 24.2 | 1.27 | 23.7 | 1.17 | 0.902 | 27.8 |
| **6** | 1 | 450 | 100.0 | 23.0 | 22.7 | 0.87 | 21.3 | 0.74 | 0.799 | 15.7 |
| **7** | 58 | 450 | 100.0 | 23.0 | 22.7 | 1.19 | 22.2 | 1.09 | 0.896 | 24.2 |
| **8** | 58 | 160 | 35.6 | 23.9 | 23.6 | 1.21 | 23.1 | 1.12 | 0.898 | 25.8 |
| **9** | 26 | 160 | 35.6 | 23.9 | 23.6 | 1.15 | 23.0 | 1.05 | 0.886 | 24.1 |
| **10** | 160 | 450 | 100.0 | 23.0 | 22.7 | 1.27 | 22.3 | 1.17 | 0.902 | 26.0 |

(#) Adopted short circuit current densities as obtained in (Cai et al. 2015). Observation: In the simplified Beer-Lambert model (Appendix B.1.1), the $J_{sc}$ is to some minute extent lower than $qG_{max}$, as used in the setup of this model (see Matlab™ program in Supplementary Material), an effect which disappears for large values of $m > 10$ in equation (8). However, as in the present modeling an $m = 4.5$ is adopted, a feeble reduction of the $J_{sc}$, in comparison to $qG_{max}$, is obtained. A meliorated model of the short circuit current density would, therefore, result in slightly higher maximal efficiencies as here presented, because of the higher resulting $J_{sc}$ values.

**Table 5**

Relative increase or decrease of different performance parameters in the comparison of several suggested design proposals.

| Row | Design improvements | $J_{SC}$ [%] | $V_{OC}$ [%] | $J_{MPP}$ [%] | $V_{MPP}$ [%] | FF [ %] | PCE [%] |
|---|---|---|---|---|---|---|---|
| 1 | Sun - Design 9 | 4.0 | 32.8 | 8.3 | 41.2 | 10.8 | 53.0 |
| 2 | Sun - Design 7 | 0.1 | 36.9 | 4.5 | 47.0 | 12.1 | 53.6 |
| 3 | Sun - Design 10 | 0.1 | 46.0 | 4.7 | 57.5 | 12.8 | 64.8 |
| 4 | Sun - Design 3 | 4.0 | 49.2 | 8.8 | 61.2 | 13.0 | 75.3 |
| 5 | Sun - Design 8 | 4.0 | 40.1 | 8.6 | 50.7 | 12.3 | 63.7 |
| 6 | Design 7 - Design 8 | 3.9 | 2.3 | 4.0 | 2.5 | 0.2 | 6.6 |
| 7 | Design 9 - Design 8 | 0.0 | 5.5 | 0.3 | 6.7 | 1.4 | 7.0 |
| 8 | Design 9 - Design 7 | -3.8 | 3.1 | -3.5 | 4.1 | 1.2 | 0.4 |
| 9 | Design 7 - Design 10 | 0.0 | 6.6 | 0.2 | 7.1 | 0.6 | 7.3 |
| 10 | Design 8 - Design 10 | -3.8 | 4.2 | -3.7 | 4.5 | 0.4 | 0.7 |
| 11 | Design 10 - Design 3 | 3.9 | 2.2 | 4.0 | 2.3 | 0.2 | 6.4 |
| 12 | Design 8 - Design 3 | 0.0 | 6.5 | 0.1 | 6.9 | 0.6 | 7.1 |
| 13 | Design 9 - Design 3 | 0.0 | 12.4 | 0.4 | 14.1 | 2.0 | 14.6 |
| 14 | Design 7 - Design 3 | 3.9 | 9.0 | 4.1 | 9.6 | 0.8 | 14.1 |

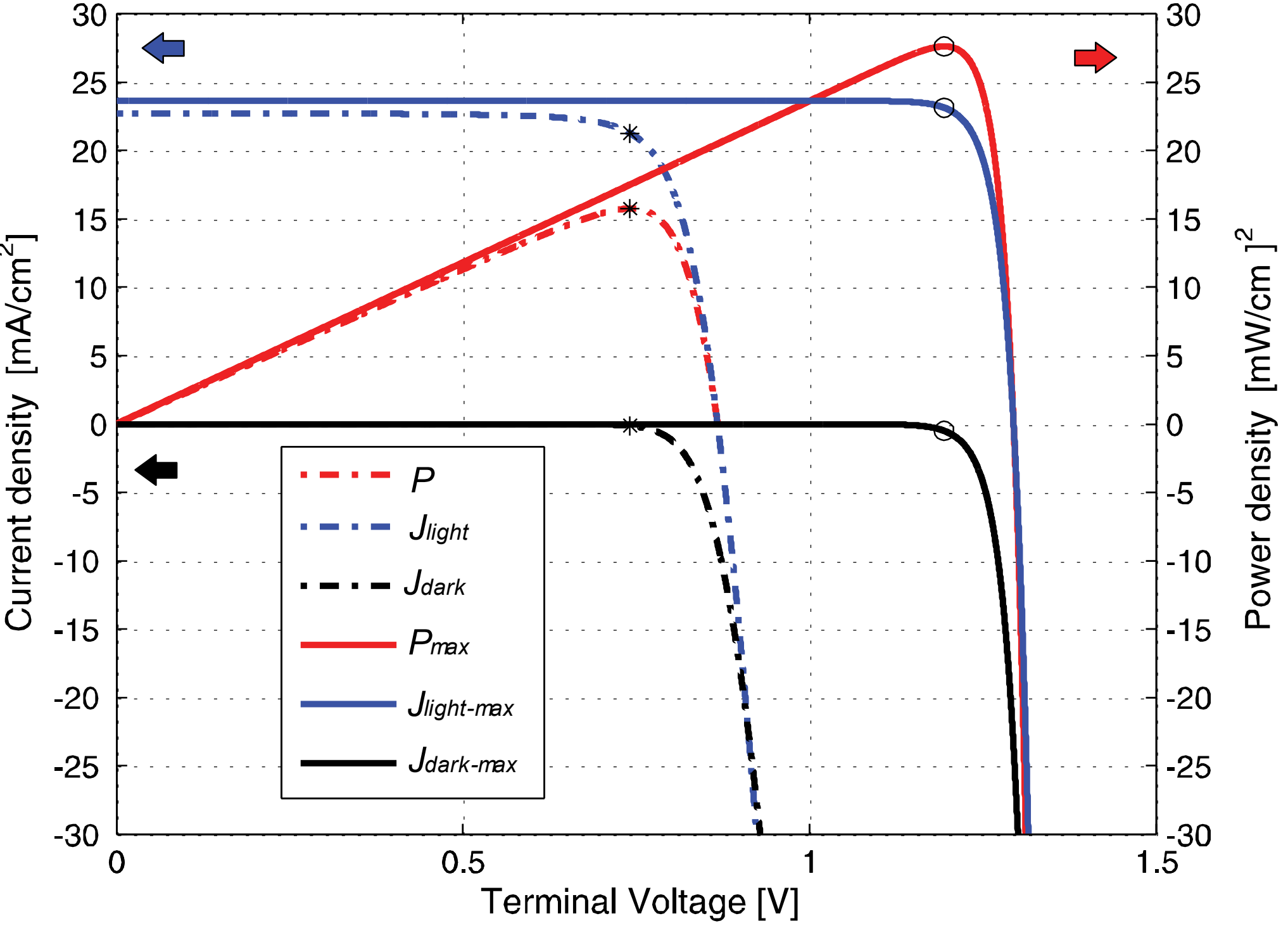

**Fig. 5.** J-V and power curves of the initial and optimized PSCs, with marked maximal power point (MPP) values: (i) curves for the solar cell configuration of the manufactured and modeled PSC in (Sun et al. 2015) ($J_{light}$, $J_{dark}$, $P$) and (ii) curves for the simulated solar cell model, as show for a multidimensional optimization, with an maximal PCE of 27.6 % for $f_B$ = 160, $t_0$ = 160, ($J_{light-max}$, $J_{dark-max}$, $P_{max}$) for AM 1.5 reference solar irradiance of 100 mW/cm². Stars and circles show the MPP operation points of these curves with MPP power densities of 15.7 mW/cm$^2$ and 27.6 mW/cm$^2$, corresponding in this normalized presentation to efficiencies of $\eta_{MPP}$ = 15.7% and 27.6% (Matlab$^{TM}$ program for the configuration of this figure as a function of the cell's material properties in Supplementary Material).

However, considering that the state-of-the-art cell is not based on light trapping (design 7); its PCE can be increased by 6.6% using light trapping with spherical nanoparticles, as based on a 4% rise of $J_{MPP}$ and a 2.5% rise of $V_{MPP}$, obtaining design 8 (Table 5 - row 6). A similar relative increase is obtained for cells with improved ($f_B \geq 58$, Fig. 3) and perfect material properties ($f_B = 160$, Table 5 - row 11). Further material property improvements in design 8 can raise the PCE by 7%, as based on a 6.9% rise of the $V_{MPP}$, obtaining cell design 3 (Table 5 - row 12). If no light trapping is considered at all, the state-of-the-art efficiency (design 7) can be increased by 7.3%, as almost only based on a 7.1% $V_{MPP}$ rise, obtaining cell design 10 (Table 5 - row 9), which presents the here considered most ideal material properties with $f_B = 160$. The efficiency of 26%, as obtained by design 10 ($f_B = 160$), can also be obtained with a much lower property improvement factor of $f_B \approx 62$, if light trapping is used in a PSC with a $t_0$ =160 nm (horizontal dotted line in Fig. 3). Finally, the highest efficiency increase (14.1 %) of design 7 is obtained, using light trapping and material property improvements combined, resulting in design 3, as based on a 9.6% increase of the $V_{MPP}$ and a 4.1% increase of the $J_{MPP}$ (Table 5 - row 14).

Fig. 5 compares the two J-V curves, and its power curves, of (i) the cell as optimized in (Sun et al., 2015) (dash-dotted curves), and (ii) the here considered, most ideal cell design. The latter was obtained by a multidimensional optimization of a cell with spherical plasmonic nanoparticles in a 160 nm thick absorber layer (Table 3 - cell design 3). The PCE of this cell is only 0.2 % lower in comparison to the highest obtained efficiency in a cell with $t_0$ = 400 nm (Fig. 3, Table 2). The improvement of the cell's material properties and the light trapping with spherical nanoparticles result in a significant PCE increase from 15.7% to 27.6%.

## 5 Discussions

Summing up the nine one-dimensional efficiency increases, as presented in Table A.1, we register a total contribution of 4.5%, for a boundary amplification factor of $f_B$ = 160. However, the combined contribution of these variables, in a multidimensional optimization, results with the maximum value of 27.6% (Table 3) in a much higher efficiency rise of 27.6% - 15.7% = 13.4%, considering the same $f_B$ factor. Consequently, because of the summed single contributions of the one-dimensional PCE increases, the efficiency raises by 4.5%, and because of further effects, a much higher growth, with the value of 13.4% – 4.5% = 8.9% can be obtained. We registered a similar behavior also in the comparison of one- and two-dimensional optimizations (Appendix, Section A.1) and assign this large additional PCE increase to the nonlinearities, as inherent to the analyzed PSC and its drift-diffusion model. As a result, higher-dimensional optimizations are much more effective, in comparison to the state-of-the-art one- and two-dimensional optimizations.

The proper efficiency growth in Fig. 3 presents also a nonlinear behavior, as the highest PCE gradients are obtained for low $f_B$ values, which correspond to small material property improvements in the hypercube space. Hence, the multidimensional efficiency optimization is most effective for cells, which efficiency is close, or lower than the state-of-the-art PCE with $f_B \approx 58$, as minor modifications in the material properties, lead to the highest efficiency gradients. Light trapping enhances this effectiveness in the multidimensional optimizations, because of the higher gradients of the related curves in Fig. 3.

The highest and second-highest values of $J_{sc}$ and $J_{MPP}$ are obtained by the light trapping cell designs 5 and 4, presenting the absorber layer thicknesses of $t_0$ = 400 and 320 nm (Table 4). Meanwhile, the largest and second-largest values of the efficiencies are obtained by cell designs 5 and 3, presenting a $t_0$ of 400 and 160 nm. In the case of $V_{0C}$ and $V_{MPP}$, the highest values are surprisingly obtained for the cell with the thinnest absorber layer (design 1), resulting inclusively to the largest FF, which means that $V_{0C}$ and $V_{MPP}$ decrease proportionally with the increase of $t_0$ (Table 3). As a result of these effects, thin absorber layers result in reasonable high efficiencies, obtaining thus the second-highest PCE with the selected favorite cell, which presents an absorber layer of $t_0$ = 160 nm (design 3).

We account the increases in $V_{0C}$ and FF, on the concomitant (i) reduction of $t_0$, and (ii) the light trapping effect, which result in its combination to high PCEs. The high $V_{0C}$ is a function of the photon recycling effect, as inherent to PSCs (Pazos-Outón et al., 2016), which increases by light

1 trapping (Sha et al., 2015 and Kirchartz et al., 2016), and results furthermore in a larger fill factor
2 (FF) (Sha et al., 2015). Thin absorber layers present lower recombination of charge carriers, as
3 discussed in (Domanski et al., 2016) and Appendix A.5. As a result, thin absorber layers increase the
4 FF and the $V_{oc}$, even without light trapping, as shown by numerical simulations in (Devi et al., 2018).
5 The lower absorber layer thicknesses can be manufactured with ultra-thin coating techniques (Liu,
6 2017), and the solvent-solvent extraction methods (Zhou et al., 2015), where the latter results in
7 smooth, and thickness adjustable absorber layers, in a large range of $t_0$ = 20…410 nm.
8 PSCs with $t_0 \approx 450$ nm do already present a low Pb content, which is similar, to the heavy-metal
9 content of state-of-the-art silicon photovoltaic modules (Green et al., 2014; Stasiulionis, 2015).
10 Fortunately, the here presented thickness reduction of the absorber layer to $t_0$ = 160 nm leads to an
11 additional reduction of the Pb residual.
12 The here used modeling of the short circuit current density presents high accuracies for the cases: (i)
13 without light trapping, and (ii) light trapping with spherical nanoparticles; because the setup of the
14 short circuit current density is based on high-resolution simulations, such as (i) the OTM modeling
15 (Sun et al., 2015), and (ii) the FDTD modeling (Cai et al., 2015) (Appendix section B.1.4).
16 Furthermore, we consider reasonable low uncertainties in the Beer-Lambert modeling (Appendix
17 B.1.1 to B.1.3). Additionally, we remark that large deviations in $\lambda_{ave}$ value result in very small
18 efficiency deviations for the same $f_B$ (Tables 1 to 3). Therefore, even under a hypothetical
19 consideration of large uncertainties in the Beer-Lambert law, and its related $\lambda_{ave}$, only small
20 uncertainties of the optimized efficiency values would be obtained.
21 The highest here achieved efficiency for single-junction PSCs of 27.76% (Table 2), is higher than (i)
22 the state-of-the-art efficiency of 24.2 % and (ii) the highest simulated efficiency of 25% (Agarwal
23 and Nair, 2014, 2015). In fact, a large set of our optimized design proposals, as presented in Fig. 3,
24 increase the state-of-the-art PCEs of manufactured and simulated solar cells. As expected, all
25 proposals present a lower efficiency than (i) the lowest value of the theoretical upper limit with PCE

= 29.9%, as considered for zero surface recombination velocities and ideal light trapping (Ren et al., 2017).

The material property improvements as selected by the multidimensional optimization algorithm are general and must not coincide with property values of manufactured PSCs, as improvements of the different properties can appear at individual scales in those cells. We used the general property improvement, as expressed by $f_B$ in order to: (i) illustrate the large efficiency advantage of multidimensional property improvements; (ii) present high- and highest-performance cell designs with low Pb content, and discuss by which techniques these cells can be manufactured; (iii) estimate and discuss the percentual improvements of several performance parameters, comparing cells in different development states, or cell designs; (iv) identify dependencies in-between model variables, as appeared between $t_0$ and $\lambda_{ave}$ (Table A.2 and Figs. 2a and 2b); (v) present a complete set of possible two-dimensional sensitivity analyses (Appendix A.1); and (vi) identified from the latter analyses that $s_f$, $s_b$, $t_0$, $\lambda_{ave}$ and $D_p$ are the most important optimization variables, in this order. As a result, we get a better understanding and provide useful knowledge for the efficiency optimization of manufactured PSCs. The presented results can be used as a roadmap, showing to which extend a PSC's performance parameters and its efficiency can be increased, by different measures and in several development states, while its Pb content is reduced. In our design proposals, we consider one, two (Table A.1), and multidimensional improvements of the cell's material properties (Fig. 3); which are combined with light trapping, and thin-film coating techniques.

Whereas material property improvements lead to the highest relative efficiency increase in cells, which PCE is less or equal to the state-of-the-art efficiency (Fig. 3, Table 5 - row 2), light trapping techniques are most efficient to improve the efficiency in cells, which PCE is higher or equal than the state-of-the-art PCE (Fig. 3, Table 5 - rows 6 and 11). While the former is almost solely based on $V_{MPP}$ increase, the latter increase both, the $J_{MPP}$ and the $V_{MPP}$. Being the state-of-the-art efficiency based on light trapping (cell design 9) or not (cell design 7), the cell's efficiency can be increased by:

(i) further material property improvements (Table 5 - row 13), and (ii) its combination with light trapping (Table 5 - row 14), resulting in both cases to a PCE increase of approximately 14%. The $\lambda_{ave}$ should be always determined as a function of $t_0$, being this in an $m$-constraint or a $f_B$-constrained relationship, where the former is simply to calculate, and the results in slightly higher efficiency values (Table 1 to 3 - last two columns), if very low reflection losses are configurable. As the remaining properties stand not in a relationship, the here presented multidimensional optimizations can be substituted by simple model simulations, in cell manufacturing and research, adopting the two-dimensional sensitivity analysis, as presented in Fig. 2f; and alternatively, using the constraint condition in equation (9). We suggest that PSC manufacturing should be based on thickness adjustable thin-film coating techniques (Zhou et al., 2015, Liu, 2017), where an ideal $t_0$ can be adjusted as a function of a configured $\lambda_{ave}$. Otherwise, adjustable light trapping techniques (Cai et al., 2015) may be used to configure an adjustable ideal $\lambda_{ave}$, for a fixed absorber layer thickness. We also like to advise that some solar cell types have an ideal built-in field (Green, 2009), as related to an ideal built-in voltage, an effect that was not observed with the present model in the constraint of $V_{bi} \leq 1.4$ V.

## 6. Conclusions

Our analyses lead to a better understanding of the cell's optimization process, and we propose high-efficiency PSC designs with ultra-thin absorber layers with a significant lower Pb content. The high PCE values for the absorber layer thickness of 160 nm, are obtained as light trapping shows similar current densities for a wide range of absorber layer thicknesses, and as $V_{0C}$ and $V_{MPP}$ increase as a function of the thickness reduction of the absorber layer. We demonstrate and discuss a large efficiency advantage, in multidimensional optimizations, and assign this effect to the nonlinear behavior of the drift-diffusion model of electrons and holes. As discussed, our optimizations result in PCE values, which appear in expected, and reasonable, ranges of the efficiency, and we consider low uncertainties for simulated design proposals. We obtained high sensitivities for low $f_B$ values, and these sensitivities improve with light trapping. Furthermore, light trapping is most efficient for high $f_B$ values, which result in the highest PCE increases. Therefore, light trapping is always advantageous. For high $f_B$ factors, as it leads to the highest PCE rises, and for low $f_B$ factors, as it not only increases the efficiency but also helps to improve the sensitivity of the further model variables. However, researchers which like to optimize PSCs with efficiencies much lower than the state-of-the-art PCE should focus principally on the improvement of the material properties, by reason of the resulting high PCE gradients in this range.

**Future works**

Further FDTD simulations for cells with shape optimized nanoparticles (Appendix sections A.2 and B.1.5) and special cell designs with high $J_{sc}$ values (Appendix section A.6 and Table A.4) should be accomplished. A protocol for the multidimensional material property measurement and control should be developed in the research with the manufactured prototypes.

## Nomenclature

|  |  |
|---|---|
| $D_n$ | Diffusion coefficient of electrons [cm²/s], |
| $D_p$ | Diffusion coefficient of holes [cm²/s], |
| $f_B$ | Boundary expansion factor [ - ], |
| $G_{AM1.5}$ | Solar irradiance with air mass 1.5 [mW/cm$^2$] |
| $G(x)$ | Generation rate of charges as a function of x [s$^{-1}$cm$^{-3}$], |
| $G_{eff}$ | Effective charge carrier generation [s$^{-1}$cm$^{-3}$], |
| $G_{max}$ | Maximal or total charge carrier generation [s$^{-1}$m$^{-2}$] |
| $J_b$ | Electron recombination current density of the back charge conduction layer at $x = t_0$ [mA/cm²], |
| $J_{dark}$ | Measurable current density in the dark [mA/cm²], |
| $J_f$ | Electron recombination current density of the front charge conduction layer at $x = 0$ [mA/cm²], |
| $J_{light}$ | The measurable current density under light exposure [mA/cm²], |
| $J_{MPP,i}$ | Maximal power point current density [mA/cm²], |
| $J_n$ | Electron current density at the back charge conduction layer [mA/cm²], |
| $J_p$ | Hole current density at the front charge conduction layer [mA/cm²], |
| $J_{sc}$ | Short circuit current density [mA/cm²], |
| $m$ | Factor which relates $t_0$ and $\lambda_{ave}$ in the Beer-Lambert law [-], |
| $P_{MPP}$ | Maximal power point power density [mW/cm$^2$], |
| $k_B$ | Boltzmann constant 1.38064852 × 10$^{-23}$ [J/K], |
| $q$ | Electric charge of an electron or hole [mAs], |
| $s_f$ | Surface recombination velocity of electrons ($s_n$) [cm/s], |
| $s_b$ | Surface recombination velocity of holes ($s_p$) [cm/s], |
| $T$ | Cell temperature [K], |
| $t_0$ | Absorber layer thickness [nm], |
| $t_{0\text{-}min}$ | Minimal absorber layer thickness [nm], |
| $t_{0\text{-}min}$ | Minimal necessary absorber layer thickness [nm], |
| $V$ | Terminal voltage of the solar cell [V], |
| $V_{bi}$ | Built-in voltage [V], |
| $V_{MPP}$ | Maximal power point voltage [V], |

$V_{oc}$ Open-circuit voltage [V],

$x = 0 \ldots t_0$ Solar irradiance penetration depth [m],

$X_j = X_1 \ldots X_9$ $X_j$ is one of the nine model variables of the PSC model,

$X_{j,me}$ A model variable extracted from the measured J-V curve, $X_j$, is one of the nine model variables of the PSC model,

$X_{j\text{-}min} \ldots X_{j,max}$ Variable expansion range for the variable j,

**Greek symbols**

$\Delta n$ Excess minority carrier concentration of electrons [$cm^{-3}$],

$\Delta p$ Excess minority carrier concentration of holes [$cm^{-3}$],

$\eta_i$ Optimized PCE for the i-th optimization iteration [%],

$\eta_{max}$ Optimized efficiency value [%],

$\lambda$ Wavelength of the solar irradiance [nm],

$\lambda_{ave}$ Average optical decay length [nm].

$\mu_n$ Drift coefficient or mobility of electrons [$cm^2V^{-1}s^{-1}$]

$\mu_p$ Drift coefficient or mobility of holes [$cm^2V^{-1}s^{-1}$]

**Indices**

$i = 1 \ldots N$ Iterations in the optimization of the efficiency,

$k = 1 \ldots M$ Iterations in the optimization of the power curve,

$j = 1 \ldots 9$ Index, which counts the nine variables,

## Declaration of Competing Interests

This manuscript does not have any conflict of interest.

## Acknowledgments

The authors thank (i) Comissão de Aperfeiçoamento de Pessoal do Nível Superior (CAPES), (ii) Fotovoltaica UFSC - Grupo de Pesquisa Estratégica em Energia Solar at Universidade Federal de Santa Catarina and (iii) Agência Nacional de Energia Elétrica (ANNEL) for their financial support.

## Appendix

The appendix and supporting material associated with this article can be found, on the online version: